\documentclass{PoS}

\usepackage{subcaption}
\usepackage{amsmath}
\usepackage{amsthm}
\usepackage{amssymb}
\usepackage{nccmath}
\usepackage{epsfig}

\title{Search for ultra-high energy photons: observing the preshower effect with gamma-ray telescopes}

\ShortTitle{Search for ultra-high energy photons through preshower effect}

\author{\speaker{Kevin Almeida Cheminant}$^{,a}$, Dariusz G\'ora$^{a}$, David E. Alvarez Castillo$^{b}$, Niraj Dhital$^{a}$, Piotr Homola$^{a}$, Pawe\l{} Jagoda$^{a,c}$, Konrad Kopa\'nski$^{a}$, Marcin Kasztelan$^{d}$, Peter Kovacs$^{e}$, Marta Marek, Vahab Nazari$^{a,e}$, Michal Nied\'zwiecki$^{f}$, Katarzyna Smelcerz$^{a,f}$, Karel Smolek$^{g}$, Jaros\l{}aw Stasielak$^{a}$, Oleksandr Sushchov$^{a}$, Krzysztof Rzecki$^{h}$, Tadeusz Wibig$^{i,d}$, Jilberto Zamora-Saa$^{l}$
\\\
E-mail: \email{kevin.almeida-cheminant@ifj.edu.pl}\\

\footnotesize
$^{a}$ Institute of Nuclear Physics Polish Academy of Sciences, Radzikowskiego 152, Cracow, Poland\\
$^{b}$ Joint Institute for Nuclear Research, Dubna, Russia  \\
$^{c}$ AGH University of Science and Technology, 30-059 Cracow, Poland\\
$^{d}$ National Centre for Nuclear Physics, Andrzeja Soltana 7, 05-400 Otwock-Swierk, Poland\\
$^{e}$ Institute for Particle and Nuclear Physics, Wigner Research Centre for Physics, Hungarian Academy of Sciences, H-1525 Budapest, Hungary \\
$^{f}$ Institute of Telecomputing, Faculty of Physics, Mathematics and Computer Science, Cracow University of Technology, Warszawska 24st 31-155 Cracow, Poland\\
$^{g}$ Institute of Experimental and Applied Physics, Czech Technical University in Prague, Czech Republic\\
$^{h}$ Cracow University of Technology, Warszawska 24st 31-155 Cracow, Poland \\
$^{i}$ University of Lodz, Faculty of Physics and Applied Informatics, Lodz, Poland\\
$^{l}$ Universidad Andres Bello, Departamento de Ciencias Fisicas, Facultad de Ciencias Exactas, Avenida Republica 498, Santiago, Chile
}

\abstract{

\scriptsize Ultra-high energy photons constitute one of the most important pieces of the astroparticle physics problems. Their
observation may provide new insight on several phenomena such as supermassive particle annihilation or the GZK effect. Because
of the absence of any significant photon identification by a leading experiments such as the Pierre Auger Observatory, we consider a
screening phenomenon called preshower effect which could efficiently affect ultra-high energy photon propagation. This effect is a
consequence of photon interactions with the geomagnetic field and results in large electromagnetic cascade of particles several
thousands kilometers above the atmosphere. This collection of particles, called cosmic-ray ensembles (CRE), may reach the
atmosphere and produce the well-known air showers. In this paper we propose to use gamma-ray telescopes to look for air showers
induced by CRE. Possible sources of ultra-high energy photons include the GZK effect and Super Heavy Dark Matter particles.
Simulations involving the preshower effect and detectors response are performed and properties of these peculiar air showers are
investigated. The use of boosted decision trees to obtain the best cosmic-ray ensemble/hadron separation, the aperture and event
rate predictions for a few models of photon production are also presented.}

\FullConference{\footnotesize 36th International Cosmic Ray Conference -ICRC2019-\\
		July 24th - August 1st, 2019\\
		Madison, WI, U.S.A.}

\begin{document}

\section{Introduction}

The mean free path of ultra-high energy (UHE) photons propagating through the Universe is constrained by their capacity to interact with electromagnetic fields and produce $e^{-}/e^{+}$ pairs. Such phenomenon reduces their astrophysical horizon down to several tens of Mpc and decreases the possibility for them to reach the Earth's atmosphere. However, as the resulting $e^{-}/e^{+}$ pairs propagate through the same electromagnetic fields, they emit lower-energy photons via bremsstrahlung radiation that may travel on longer distances in the form of cosmic-ray ensembles (CRE). These CREs may reach the top of the atmosphere with various spatial and time distributions that directly depend on the distance $d_{int}$ between the UHE photon conversion altitude and the atmosphere, and also on the characteristics of the local electromagnetic field. The extensive air showers (EAS) subsequently produced might be detected on ground by various experiments, including gamma-ray telescopes.

\begin{figure*}[!t]
  \centering
  \includegraphics[width=8cm]{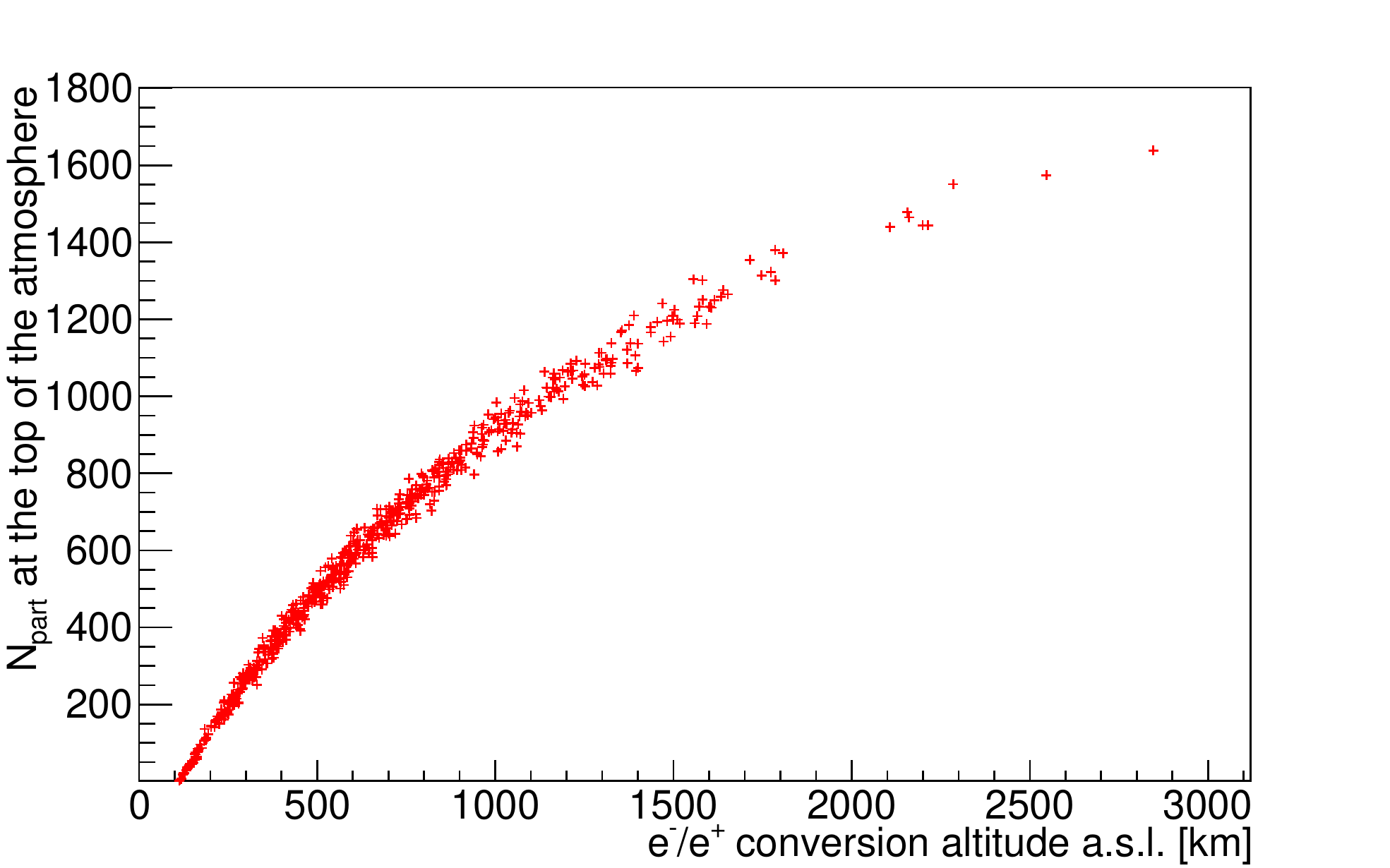}
  \caption{Number of secondary particles reaching the top of the atmosphere as a function of the altitude at which a 40 EeV photon converts into a $e^{-}/e^{+}$ pair.}
  \label{fig:altitude}
\end{figure*}

In this paper, we propose to study the case of UHE photons emitted by a diffuse source and which would interact in the geomagnetic field to produce a cascade of electromagnetic particles just above the Earth's atmosphere called \textit{preshower} \cite{mcbreen81}. The distance $d_{int}$ being relatively small (up to a few thousands of km), the CREs reaching the top of the atmosphere are contained within up to a few square centimeters. The number of particles reaching the top of the atmosphere depends directly on the altitude at which the UHE photon converts into a $e^{-}/e^{+}$ pair as shown in Fig.\ref{fig:altitude}. The greater the altitude, the more bremsstrahlung photon the pair can radiate. We investigate the possibility to observe and identify the EAS produced by preshowers with the next-generation gamma-ray telescope array in the northern hemisphere, Cherenkov Telescope Array (CTA) - La Palma \cite{cta11}. We argue that by adopting an observation mode at high zenith angles and by making use of Boosted Decision Trees (BDT) classification method, these preshowers can be discriminated from the cosmic-ray (CR) dominated background and provide new insights on the existence and nature of UHE photons.

In the first section, we describe the simulation chain used to generate preshowers and CR initiated EAS as well as to reproduce CTA cameras response before briefly introducing the BDT method. Then, we present the results of the binary classification and calculate CTA-La Palma effective area as well as the number of expected preshower events from different UHE photon production scenarios. In the conclusion, we open the discussion regarding the possibility to observe such events within the framework of the Cosmic-Ray Extremely Distributed Observatory (CREDO) experiment \cite{credo}.

\section{Method}

\subsection{Simulations} \label{Sim}
The preshower effect is simulated via the \textit{PRESHOWER} algorithm \cite{homola05}. It performs the calculation of the $e^{-}/e^{+}$ conversion probability as a UHE photon propagates through the geomagnetic field (IGRF model \cite{igrf}), and includes the deviation of the pair's trajectory and its emission of braking radiation. More details regarding the bremsstrahlung radiation energy spectrum and the interaction probabilities are given in \cite{homola05}. 

The CORSIKA 6.990 software \cite{heck98} is used to simulate the development of EAS in the atmosphere. It describes the particle cascade using different interaction models. In our case, QGSJETII-03 \cite{ostapchenko06} and URQMD \cite{bass98} have been selected to account for high and low energy hadronic interactions, respectively. The VIEWCONE option allows us to simulate diffuse sources by randomly selecting a shower direction within a cone of opening angle $\alpha$. It has been set to $\alpha = 5^{\circ}$ for both preshower and CR background simulations. The CSCAT option was also chosen to investigate the trigger efficiency by randomizing the impact location of the shower core within a circle of radius $R_{imp}$ in the plane perpendicular to the shower direction. In our simulations, $R_{imp}$ was set to 1300 m. Finally, as gamma-ray telescopes detect the Cherenkov light emitted by the charged particles within the EAS, the CERENKOV option was activated to account for such emissions. 

The northern site of CTA will be composed of 15 medium-sized telescopes with a field of view (FoV) of $8^{\circ}$ and 4 large-sized telescopes with a $5^{\circ}$ FoV. In order to simulate the cameras response, we pipe the outcome of CORSIKA into the \textit{sim\_telarray} package with the \textit{production-I} settings \cite{bernlohr08} (all parameters regarding the trigger system and the electronics are left to default values and only events triggering 2 telescopes or more are kept).

In this study, we choose to look at a UHE photon with energy $E=40$ EeV which constitutes a good compromise between $e^{-}/e^{+}$ conversion rate and the computational time required to run the whole simulation chain. In fact, such conversion rate being directly dependent of the geomagnetic field local characteristics, the direction of arrival highly influences such rate. In the case of La Palma location, UHE photons are more likely to convert if coming from the geomagnetic North ($\phi = 180^{\circ}$ in CORSIKA reference frame) where the electromagnetic field is stronger \cite{almeida17}. Also, high zenith angles directions (here, set to $\theta = 80^{\circ}$) are considered for two reasons. First of all, the same conversion rate is higher for UHE photons that have to travel through a larger section of the geomagnetic field. More importanly, the images formed on the cameras are mostly coming from the Cherenkov light emitted by muons when observing at high zenith angles. Indeed, the other components (hadronic and electromagnetic) of the EAS are significantly absorbed as it develops through a thicker atmospheric layer while muons tend to propagate on much larger distances. The muon component being directly related to the nature of the primary particle, it was shown that observing at high-zenith angles allowed to isolate this component and to retrieve a good gamma-hadron separation \cite{neronov16}.

Regarding the CR background simulations, we limited ourselves to proton primaries with an energy range between $10^{4}$ GeV, value below which very few air showers make it to the ground detectors in the nearly-horizontal direction and $3\times10^{6}$ GeV, value above which the number of expected CR falls below 1 event for a typical observation time of 30 hours and when considering the maximum impact distance $R_{imp}=1300$ m. The energy spectrum is simulated according to the power-law $\propto E^{-2.7}$.

\begin{figure*}[!t]
  \centering
  \includegraphics[width=13.7cm]{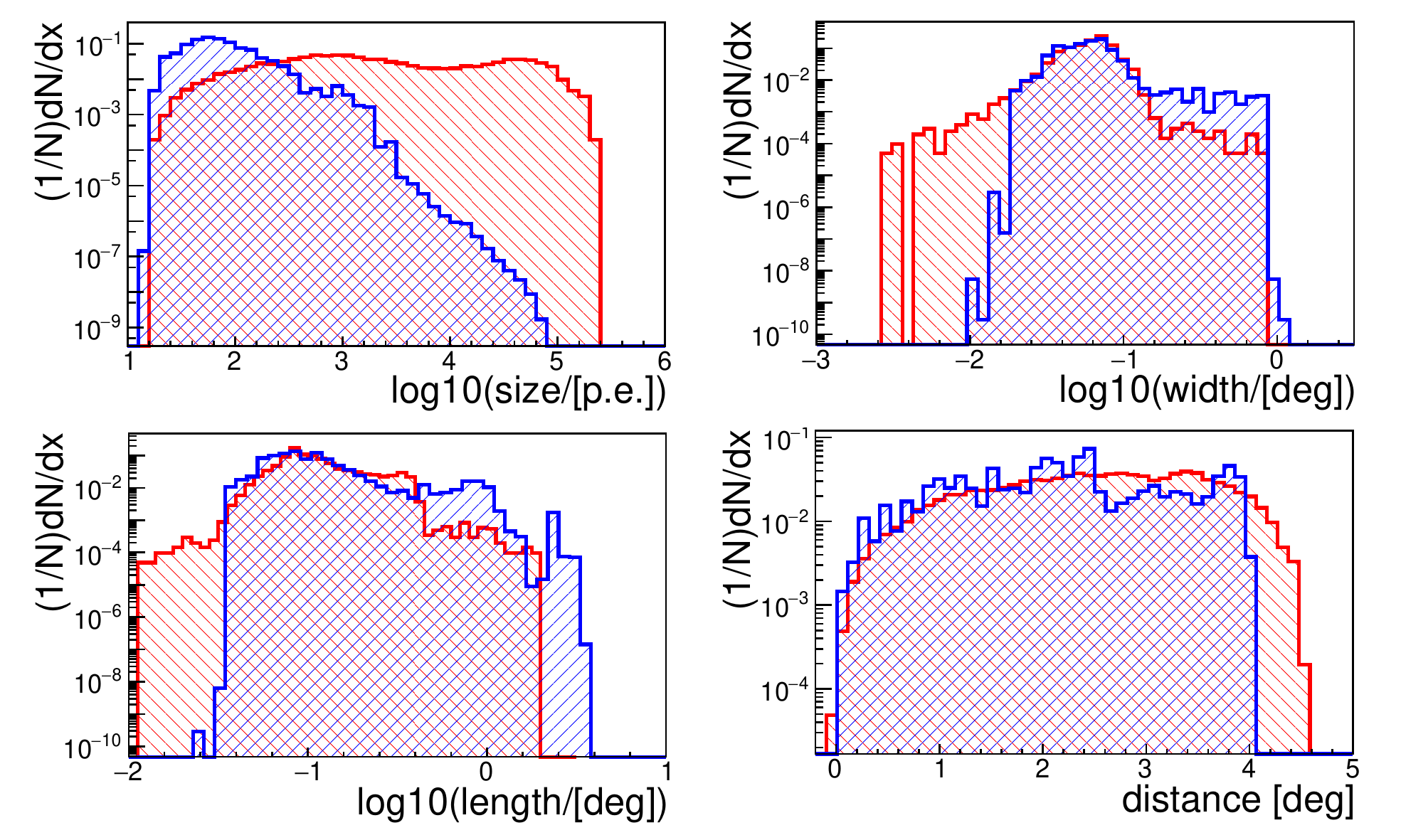}
  \caption{Distribution of Hillas parameters for simulated preshower effect (red) and CR background (blue).}
  \label{fig:training_variables}
\end{figure*}%

\subsection{Multivariate analysis}

The Cherenkov light emitted by the EAS is recorded by the telescopes cameras and the images formed are characterized by a set of parameters called Hillas parameters \cite{hillas85}. The size (number of photo-electrons), the width, the length and the distance (between camera center and image center) parameters are used to feed a BDT classifier provided by the TMVA package \cite{tmva}. The principle is to obtain two distributions (one for preshowers and one for CR background) of 'BDT scores' as separatated as possible, based on the distributions of the four parameters listed previously. A set of so-called \textit{training} events is used to train the classifier while another set of \textit{testing} events is used to evaluate how good is the trained classifier at correctly identifying independent preshower and CR background events. The details of the BDT method in the context of gamma-ray astronomy can be found in papers published by HESS \cite{ohm09} and MAGIC \cite{albert08} that show its efficiency in this framework. In our study, the parameters of the method were left to the default values provided by TMVA with very little tuning. 

\section{Results}

\subsection{Classification}

The distributions shown in Fig.\ref{fig:training_variables} are those of the Hillas parameters for the preshowers (red) and the CR background (blue). The \textit{size} distributions show that overall, preshower-induced images tend to be brighter than CR background ones. It is mostly due to the larger primary energy and to the fact that preshower-induced EAS reach their maximum closer to the ground. The \textit{length} and \textit{parameters} describe the geometrical dimensions of the images. Because CR background EAS are more irregular, these parameters tend to be shifted towards larger values. Finally, the \textit{distance} parameter being directly connected to the angle between the EAS axis and the pointing direction of the camera, the two distributions are very similar as a direct consequence of the same viewing cone angle $\alpha = 5^{\circ}$ used to simulate the diffuse nature of both primaries.

\begin{figure*}[!t]
  \begin{subfigure}{.5\textwidth}
    \centering
    \includegraphics[width=7.5cm]{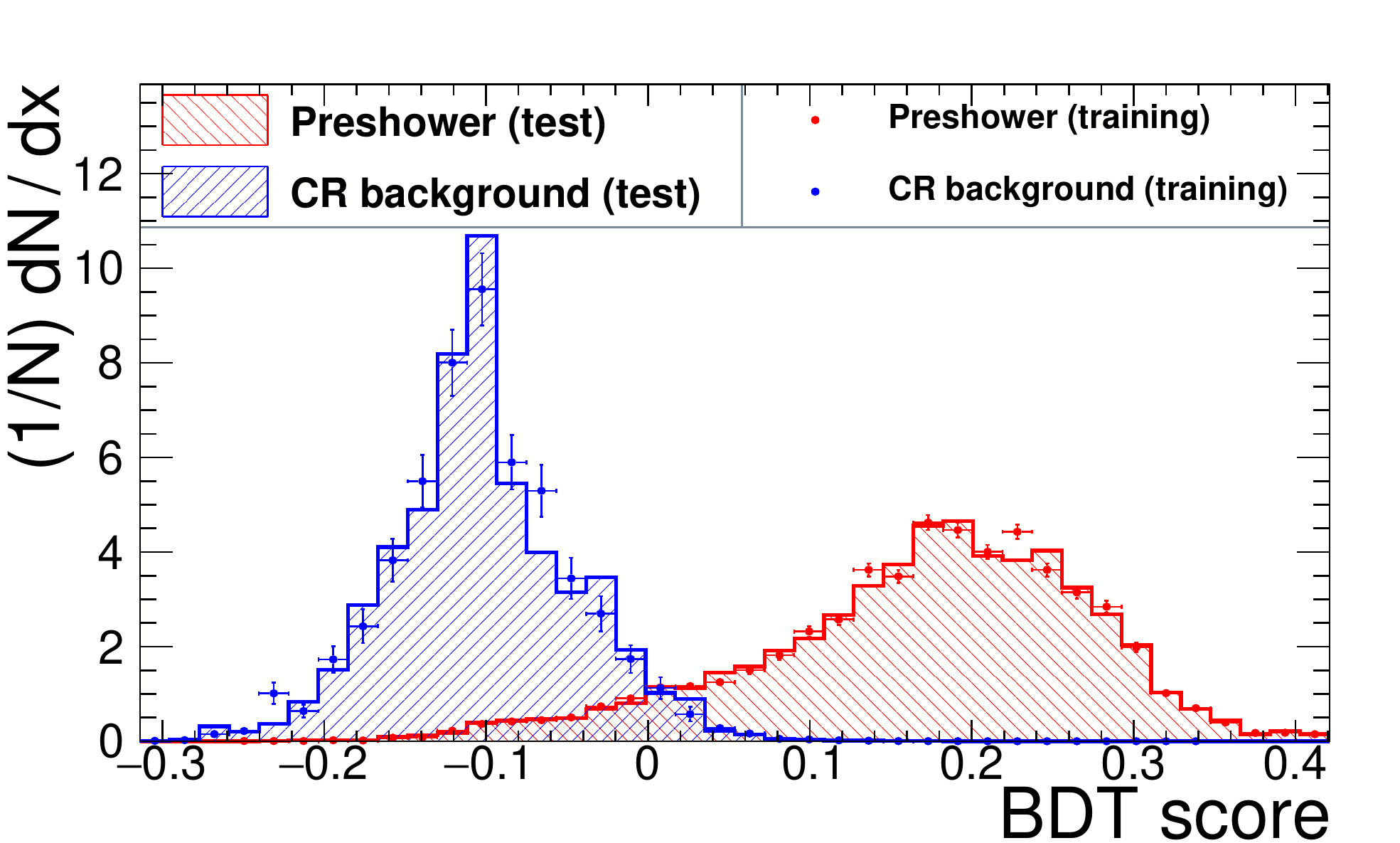}
  \end{subfigure}
  \begin{subfigure}{.5\textwidth}
    \centering
    \includegraphics[width=7.5cm]{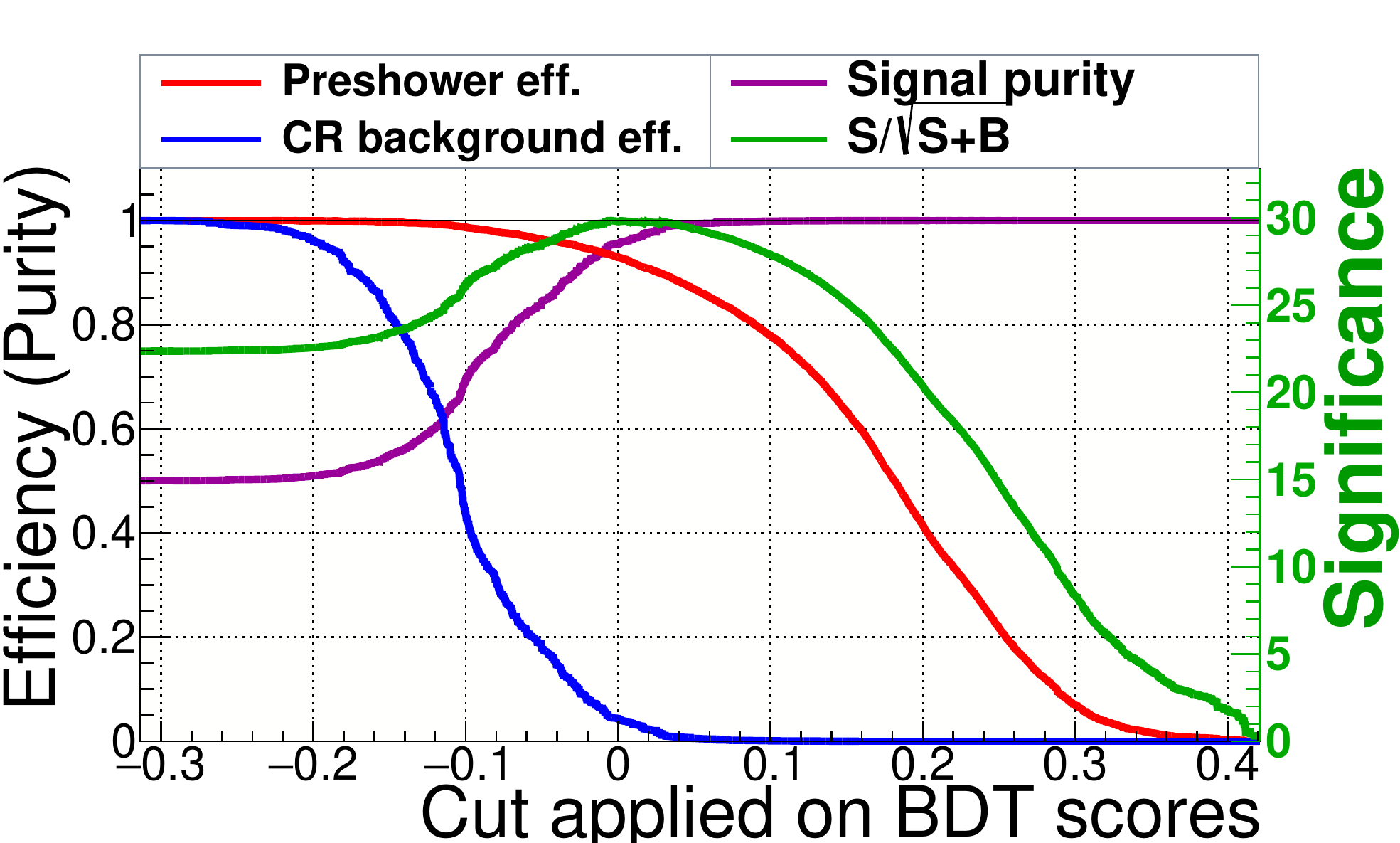}
  \end{subfigure}
  \caption{\textit{Left panel}: BDT score distributions of the testing and training samples of CR background and preshowers. \textit{Righ panel}: Efficiencies, purity of preshower sample and significance of the classification as a function of the cut value applied to the distributions shown on the left side.}
  \label{fig:BDTdist}
\end{figure*}

The distributions described above are fed into the BDT method of the TMVA package in order to investigate if preshowers and CR background events can be effectively discriminated. By assigning a score to each event contained in the training sample, two BDT score distributions are obtained as shown in Fig.\ref{fig:BDTdist} (left). The more these two distributions are separated, the better the classifier is at correctly identifying events that were not used to train the method (test sample). To investigate the efficiency of the method, one can apply different cut values to the test sample distributions and see how the preshower and CR background efficiencies evolve. Such analysis is automatically performed by TMVA and is shown in Fig.\ref{fig:BDTdist} (right). To identify which cut value is best, we decided to rely on the significance calculation provided by TMVA. For different cut values, one can calculate and obtain the best $\sigma$ coefficient by looking for the maximum of Eq.\ref{eqn:sig} in order to evaluate the quality of the classification of the training sample:

\begin{ceqn}
  \begin{align}
    \sigma =  \frac{S}{\sqrt{S+B}} \label{eqn:sig}
  \end{align}
\end{ceqn}

where S is the preshower efficiency multiplied by the number of true positives (preshower events classified as preshower) and B is the background efficiency multiplied by the number of false negatives (CR background events classified as preshower). Using this method, the maximum $\sigma = 29.85$ is found at a cut value of \textbf{-0.01}. For such cut, the preshower efficiency is $\epsilon_{preshw}=0.926$ (meaning that 92.6\% of preshower events are correctly identified) and the CR background contamination is $\epsilon_{CR}=0.038$ (meaning that 3.8\% of CR background events will be identified as preshower events). Of course, some other criteria can be used to identify which cut value is best and compromises must be made. One could also be looking for a cut value that produces a background free sample ($\epsilon_{CR}=0$, in which case the signal efficiency tends to be lower) or that correctly identifies 100\% of the signal events ($\epsilon_{preshw}=1$, in which case a higher background contamination is expected).

Another preshower scenario was also studied, similar to the one described above except that the viewing cone angle $\alpha$ was set to $0^{\circ}$ in order to simulate a point source. The Hillas distributions are in fact very similar to those of the diffuse source scenario except for the distance since the angle between the shower axis and the pointing direction of the telescopes is defined by a constant offset of $0.5^{\circ}$. Therefore, the typical values of this parameter for a point source scenario are much smaller. Preshower efficiency and CR background contamination were found at $\epsilon_{preshw}=0.982$ and $\epsilon_{CR}=0.022$, respectively. 

\subsection{Aperture and expected number of events} \label{ap}

Using the calculated efficiencies, we can investigate the aperture of CTA-La Palma for both types of preshower scenario. The aperture is defined as:

\begin{ceqn}
  \begin{align}
    A_{eff}(E,\theta,\phi,\epsilon)=\pi R_{imp}^{2}\Omega\frac{N_{trig}(E,\theta,\phi,\epsilon)}{N_{sim}} \label{eqn:effective_area}
  \end{align}
\end{ceqn}

where $R_{imp}=1300$ m is the impact distance defined previously, $N_{trig}(E,\theta,\phi,\epsilon)$ is the number of events that have triggered the telescope array and that have been correctly assigned to their class (preshower or CR background), $N_{sim}$ is the number of events that where simulated and $\Omega$ is the solid angle in steradian ($\Omega=2\pi(1-cos(\alpha))$ for diffuse sources, where $\alpha=5^{\circ}$ is the viewing cone angle previously defined).
For the diffuse preshower source, $N_{trig}(E,\theta,\phi,\epsilon)/N_{sim}=0.42$ such that we obtain $A_{eff}=0.05\mbox{ }\mathrm{km^{2}}$.
Once the aperture for the preshower obtained, one can calculate how many events can be expected to be observed according to different UHE photon production models as well as investigate what should be the UHE photon flux in order to obtain at least one event. This event rate can be experessed as:

\begin{ceqn}
  \begin{align}
    N_{preshw}(40 EeV)=\phi_{UHE-\gamma}(40\mbox{ }\mathrm{EeV})\epsilon_{conv}\Delta tA_{eff} \label{eqn:N_e}
  \end{align}
\end{ceqn}

where $\epsilon_{conv}=0.67$ is the probability for a UHE photon to produce a preshower with the energy and direction defined in Sec.\ref{Sim} , $\Delta t = 30$h is the chosen time of observation and $A_{eff}$ is the aperture previously calculated. The integral photon flux $\phi_{UHE-\gamma}$ (in $\mathrm{km^{-2}year^{-1}sr^{-1}}$) is obtained from various UHE photon production models such as Super-Heavy Dark Matter (SHDM) model \cite{shdm} and GZK photon model \cite{gzk}, and from Pierre Auger Observatory limits on the UHE photon flux \cite{auger} at 40 EeV. Results are summarized in Tab.\ref{tab:N_E} and show that very few events would actually be detected, mainly due to the very low expected UHE photon flux from different models. In fact, we may also calculate the UHE photon flux needed in order to obtain at least one event in a 30 hours time period by setting $N_{preshw}(40\mbox{ }\mathrm{EeV}) = 1$. Such constrain leads to a necessary UHE photon flux $\phi_{UHE-\gamma}(40\mbox{ }\mathrm{EeV}) = 8.73\times10^{3}\mbox{ }\mathrm{km^{-2}year^{-1}sr^{-1}}$.

Similar calculations were performed for the point source scenario ($\alpha=0^{\circ}$ and $\Omega=1)$. In this case, we obtain $N_{trig}(E,\theta,\phi,\epsilon_{BDT})/N_{sim}=0.64$ and $A_{eff}=3.40\mbox{ }\mathrm{km^{2}}$. The results regarding the expected number of events in that particular case are summarized in Tab.\ref{tab:N_E}.

\begin{table*}[!t]
\centering
\begin{tabular}{clll}
&\multicolumn{1}{c}{SHDM} & \multicolumn{1}{c}{AUGER limits} & \multicolumn{1}{c}{GZK} \\ \hline \hline
$\alpha=0^{\circ}$ (point source)&\multicolumn{1}{c}{$1.03\times10^{-4}$}&\multicolumn{1}{c}{$3.82\times10^{-6}$}&\multicolumn{1}{c}{$3.50\times10^{-6}$}\\
$\alpha=5^{\circ}$ (diffuse source)&\multicolumn{1}{c}{$1.49\times10^{-6}$}&\multicolumn{1}{c}{$5.61\times10^{-8}$}&\multicolumn{1}{c}{$5.15\times10^{-8}$}\\
\end{tabular}
\caption{Expected number of preshowers for different UHE photon production scenarios and for aperture and time of observation given in Sec.\ref{ap}. These values are obtained using Eq.\ref{eqn:N_e}.}
\label{tab:N_E}
\end{table*}

\section{Conclusion}

The lack of conclusive observation of photons in the EeV regime \cite{auger2} and beyond constitutes a serious line of research for the whole astroparticle community. Aside from putting severe constraints on UHE photon production models, it is also deeply connected to the mystery lying behind UHE cosmic-rays and their origin. However, such non-observation must be carefully evaluated. First of all, the current EAS reconstruction technics may not be yet capable of distinguishing UHE photons and preshower-induced EAS from the hadronic ones. Secondly, screening mechanisms, such as the preshower effect demonstrated here, may significantly reduce the astrophysical horizon of UHE photons. These mechanisms may also occur far away from Earth and change the imprint of the preshower effect on the Earth's atmosphere by showing different time and spatial distributions for the secondary particles. Such phenomena might be detectable via a large network of astroparticle detectors and a corresponding strategy has been adopted by the CREDO collaboration. 

In this paper, we have shown that preshowers occuring in the geomagnetic field may be effectively distinguished from the omnipresent CR background using the future CTA-La Palma by adopting a nearly-horizontal observation mode and a multivariate analysis on the recorded images. However, one must keep in mind that the current models of UHE production are predicting extremely low flux of UHE photons, which may be counterparted by larger networks and/or longer time of observations. To obtain at least one event for a typical 30 hours observation time, one would need a photon flux much larger than the one predicted by the most optimistic models such as SHDM model. However, choosing the right cut values for theat the multivariate analysis stage would allow for a completely background free observations.

\acknowledgments

This research has been supported in part by PLGrid Infrastructure. We warmly thank the staff at ACC Cyfronet AGH-UST, for their always helpful supercomputing support.

\end{document}